\documentclass[11pt]{article}
\usepackage{graphicx}
\graphicspath{{C:/Users/xinwei/Documents/6EPMapping/gunterPaper/}}

\columnwidth\textwidth

\def\al{\alpha}

\def\ro{\varrho}

\def\d{\partial}
\def\=d{\,{\buildrel\rm def\over =}\,}

\def\sqr#1#2{{\vcenter{\vbox{\hrule height.#2pt\hbox{\vrule width.
#2pt height#1pt \kern#1pt \vrule width.#2pt}\hrule height.#2pt}}}}

\def\sq{\hbox{\rlap{$\sqcap$}$\sqcup$}}

\def\te{\vartheta}
\def\e{{\rm em}}
\def\o{{\rm obs}}
\def\B{\Bigl}

\def\diag{{\rm diag}}
\def\crit{{\rm crit}}
\def\n{\nabla}

\begin{document}

\title{Perturbation theory in nonstandard cosmology}
\author{G\"unter Scharf
\footnote{e-mail: scharf@physik.uzh.ch}
\\ Physics Institute, 
\\ University of Z\"urich, 
\\ Winterthurerstr. 190 , CH-8057 Z\"urich, Switzerland}

\date{}

\maketitle\vskip 3cm

\begin{abstract} 

We study inhomogeneous perturbations away from the strongly homogeneous background cosmology previously studied. The problem is greatly simplified by using the mapping on the inner Schwarzschild solution. The resulting linear perturbation equations can be solved by power series. The solution leads to a finite energy density, but the pressure must be zero.


\end{abstract}

\newpage

\section{Introduction}

The isotropy of CMB clearly shows that the cosmic gravitational field is spherically symmetric up to small anisotropies. However the visible matter is certainly anisotropic, furthermore its density is small compared to the critical density. In this situation a {\it realistic} description should start from a spherically symmetric vacuum solution of Einstein's equation, followed by the analysis of its anisotropic perturbations. The latter should explain both the CMB anisotropies and the finite matter density.
The question then is, which vacuum solution should be chosen as background. This problem was solved in 
a previous paper [1]. If one assumes a one-to-one correspondence between comoving coordinates and the cosmic rest frame, then there is essentially only one spherically symmetric solution of Einstein's equations simultaneously in both systems. It is the homogeneous Datt or Kantowski-Sachs solution ([3], p.110) which in comoving coordinates has a line element of the form
$$ds^2=dt^2-X(t)^2dr^2-Y(t)^2(d\te^2+\sin^2\te d\phi^2).\eqno(1.1)$$
Needless to say that we prefer the $+---$ metric because considering spin 2 gauge theories on the same basis as spin 1 [2], we cannot accept equations as $p^2=-m^2$ which follow with the signature $-+++$. In contrast to standard FLRW cosmology there is no factor $r^2$ in front of $Y(t)$, therefore we call the metric (1.1) strongly homogeneous, because there is no $r$-dependence in the metric functions. 

There is some confusion in the literature about the use of the notions ``homogeneous'' and ``isotropic''. For example Kantowski-Sachs call the solution (1.1) anisotropic although it is spherically symmetric and the angular variables can be separated in the perturbation theory. The notion ``homogeneous'' too has various different meanings [3]. In addition, it is difficult to select the correct name for (1.1) from the authors Schwarzschild, Lema\^itre, Datt, Ruban, Kantowski-Sachs etc. So we shall simply say nonstandard background in the following.

In the standard FLRW model one needs high-density hypothetical sources of the gravitational field, dark matter, dark energy or cosmological constant and an inflaton field, in order to get agreement with observations. Since none of these sources has been {\it directly} observed, non-standard cosmology based on (1.1) starts without sources from a vacuum solution which is consistent with the magnitude-redshift data. In fact it was shown in [1] sect.7 that one can get an excellent representation of the current Hubble data by means of the vacuum solution corresponding to (1.1).  Matter and radiation are considered as small inhomogeneous perturbations on the nonstandard background (1.1). The necessary perturbation theory is developed in this paper.

It is known ([3], p.386) that the nonstandard solution can be mapped on the inner Schwarzschild solution by exchanging time with the radial coordinate $r$. The mapping then is a consequence of Birkhoff's theorem. Do we really live inside of a black hole ? Not at all. The mapping is a purely mathematical operation by means of suitable (Schwarzschild) coordinates which have no physical meaning in the cosmological context. For the physical interpretation of results and comparison with observations one must always transform back to the comoving coordinates used in (1.1) or to the universal cosmic rest frame [1]. But to solve the equations of perturbation theory the mapping to the Schwarzschild solution is very useful.
Frank J. Zerilli has developed the perturbation theory of the outer Schwarzschild solution in his Ph.D thesis of 1969 [4]. With the appropriate sign changes his equations can be taken over to our situation. Then transforming back to comoving coordinates gives us the results for our Universe.
 
The paper is organized as follows. In the next section we discuss the mapping between the nonstandard solution and the inner Schwarzschild solution. In sect.3 this correspondence is used to obtain the equations of first order perturbation theory. This is a direct application of the PhD thesis of  F.J.Zerilli [4]. Zerilli has reduced the resulting linear differential equations to a wave equation of Schr\"odinger type which, however, cannot be solved analytically. Instead we transform in sect.4 to appropriate coordinates which are linearly related to the redshift. Then the perturbative equations are solved by power series. In sect.5 we briefly discuss the possible applications of the results in cosmology.

In a second paper the first order perturbation theory in the cosmological context has been rederived in detail, because the calculations are lengthy and complicated and should be available in the literature. As required by the arXiv-moderation this paper is added here as Part 2 (sections 6-11).

\section{Mapping on the inner Schwarzschild solution}

Instead of using Bondi's symbol $Y(t)$ in (1.1) as before [5], we shall now write
$$ds^2=dt^2-X(t)^2dr^2-R(t)^2(d\te^2+\sin^2\te d\phi^2)\eqno(2.1)$$
because we need the symbol $Y(\te,\phi)$ for the spherical harmonics. A second reason for this is that we are going to consider $R(t)$ as a new (radial !) coordinate. In [1] we have found the following parametric representation for $R(t)$ and $X(t)$ (equ.(5.14) and (5.18))
$$R(t)=T_L\sin^2w,\eqno(2.2)$$
$$X(t)=Q\cot w+P(1-w\cot w)\eqno(2.3)$$ 
where $P$ and $Q$ are constants of integration. The comoving time is given by (equ.(5.8) in [1])
$$t=T_L(w-\sin w\cos w)\eqno(2.4)$$
here $T_L$ determines the lifetime of the Universe. In addition we know from [1] equ.(5.25) that $X(t)$ is proportional to
$$\dot R(t)=2T_L\sin w\cos w{dw\over dt}.\eqno(2.5)$$
Substituting
$${dt\over dw}=2T_L\sin^2w$$
herein we obtain
$$\dot R(t)={\cos w\over\sin w}=\cot w\eqno(2.6)$$
hence the constant $P$ in (2.3) is zero:
$$X(t)=Q\cot w.\eqno(2.7)$$

For the physical discussion we need the radial null geodesics given by the wave vector $k^\mu=(1/X,-1/X^2,0,0)$. Then the redshift is equal to
$$1+z={X_\o\over X_\e}={\cot w_\o\over\cot w_\e}\eqno(2.8)$$
where $\e$ and $\o$ refer to the time of emission and observation, respectively. The Big Bang corresponds to $z=\infty$, that means $w_\e=\pi/2$, and $\pi/2<w_\o<\pi$, because $z$ (2.8) must be positive. This variation of $w$ is different from the one in the cosmic rest frame as discussed in [1]. From
$$dz={\cot w_\o\over\cot^2w}{dw\over\sin^2w}={\cot w_\o\over\cos^2w}{dt\over 2T_L\sin^2w}\eqno(2.9)$$
we identify the Hubble constant
$${dz\over dt}\B\vert_{z=0}=-H_0={(1+\cot^2w_\o)^2\over 2T_L\cot w_\o}.\eqno(2.10)$$

To calculate the radial distance we integrate
$${dr\over dz}={dr\over dt}{dt\over dz}={2T_L\cot ^2w\over X\cot w_\o}={2T_L(1+z)^3\over [(1+z)^2+\cot^2w_\o]^2}$$
from $z=0$ to $z>0$. With the new variable of integration $x=1/(1+z)$ we get
$$r(z)=2T_L\int\limits_{1/(1+z)}^1{dx\over x(1+\cot^2w_\o x^2)^2}.\eqno(2.11)$$
For comparison with the result in the cosmic rest frame [1] equ.(6.16) we introduce the parameter
$$\al={1\over\vert\cot w_\o\vert}\eqno(2.12)$$ 
which is also equal to the local light speed $c_0=dr/dt=1/\vert X\vert$. Using the Hubble constant (2.10) we finally obtain
$$r(z)={c_0\over H_0}(1+\al^2)^2\int\limits_{1/(1+z)}^1{dx\over x(\al^2+x^2)^2}\eqno(2.13)$$
in agreement with the result in the cosmic rest frame. Then the same luminosity distance and magnitude - redshift relation come out. The quantity $T_L$is of the order of the Hubble time because
$$H_0={(\al^2+1)^2\over 2T_L\al^3}.\eqno(2.14)$$
The parameter $\al^2$ is determined by the Hubble data and is equal to 6.71 in our Universe [1].

To map the above solution to the Schwarzschild solution we choose $R$ as a new coordinate. Then we have
$$\sin w=\sqrt{{R\over T_L}}$$ 
so that
$$X^2=Q^2{T_L-R\over R}\eqno(2.15)$$
and
$$dt=2T_L\sin^2w\,dw=2R\,dw=\tan w\,dR.$$
Then $ds^2$ (2.1) assumes the following form
$$ds^2=\B({R\over T_L-R}\B)dR^2-Q^2{T_L-R\over R}dr^2-R^2(d\te^2+\sin^2\te d\phi^2).\eqno(2.16)$$
The integration constant $Q$ was not fixed in (2.7). Now we must take $Q^2=1$ to have a solution of Einstein's equations and this is the inner Schwarzschild solution because $R<T_L$ by (2.2), the letter $Q$ is then free for later use. However the comoving radial coordinate $r$ now sits at the place of the time coordinate, we shall write
$$S=r\eqno(2.17)$$
in the following, so that $(S, R)$ are our inner Schwarzschild coordinates:
$$ds^2=g^0_{\mu\nu}dx^\mu dx^\nu =\B({R\over T_L-R}\B)dR^2-{T_L-R\over R}dS^2-R^2(d\te^2+\sin^2\te d\phi^2).\eqno(2.18)$$
This is our background metric in the new coordinates. The components of the Jacobian are equal to
$${\d t\over\d R}=-\sqrt{{R\over T_L-R}},\quad{\d r\over \d S}=1\eqno(2.19)$$
and otherwise zero. The minus sign follows from (2.6) because $\pi/2<w<\pi$. Our lifetime constant $T_L$ is equal to $2m$ where $m$ is the mass of the black hole. The Big Bang $w=\pi/2$ corresponds to the horizon $R=T_L$ whereas $w=\pi$ is the true singularity $R=0$ of the Schwarzschild solution.

Since the Schwarzschild background (2.18) is spherically symmetric, the first order perturbation $h_{\mu\nu}(S,R,\te,\phi)$ is expanded in tensor harmonics. The most explicit expressions for the 10 harmonics are given by Zerilli [4] in his Appendix A. Only the notation of these harmonics has been changed in later work [6]. For this reason we give the connection between Zerilli's notation (in brackets) and the one of Gerlach and Sengupta [6]:
$$(a_{LM}^{(0)},\, a_{LM}^{(1)},\, a_{LM})=Y(\te,\phi)\eqno(2.20)$$
$$(b_{LM}^{(0)},\, b_{LM})=Y,_a\eqno(2.21)$$
$$(c_{LM}^{(0)},\,c_{LM})=S_a\eqno(2.22)$$
$$(d_{LM})=S_{a;b}\eqno(2.23)$$
$$(g_{LM})=Y\gamma_{ab}\eqno(2.24)$$
$$(f_{LM})=Y,_{a;b}.\eqno(2.25)$$
The harmonics (2.22-23) have odd parity and are called axial or magnetic, the other seven have even parity and are called polar or electric. These all are $4\times 4$ symmetric tensors which are orthonormal on the 2-sphere. For later use we write down the following four harmonics 
$$a_{LM}^{(0)}=\pmatrix{Y_{LM}&0&0&0\cr 0&0&0&0\cr 0&0&0&0\cr 0&0&0&0\cr}\quad a_{LM}=\pmatrix{0&0&0&0\cr 0&Y_{LM}&0&0\cr 0&0&0&0\cr 0&0&0&0\cr}\eqno(2.26)$$
$$a_{LM}^{(1)}={i\over\sqrt{2}}\pmatrix{0&Y_{LM}&0&0\cr Y_{LM}&0&0&0\cr 0&0&0&0\cr 0&0&0&0\cr}\quad g_{LM}={R^2\over\sqrt{2}}\pmatrix{0&0&0&0\cr 0&0&0&0\cr 0&0&1&0\cr 0&0&0&\sin^2\te\cr}Y_{LM}.\eqno(2.27)$$

As perturbing energy-momentum tensor we consider a perfect fluid with
$$t_{\mu\nu}=(\ro+p)u_\mu u_\nu-pg_{\mu\nu}\eqno(2.28)$$
where $\ro$ and $p$ are density and pressure of the matter and $u_\mu$ its 4-velocity, the latter is $u^0_\mu=(1,0,0,0)$ in comoving coordinates. The first order perturbation of the vacuum is given by
$$\delta t_{\mu\nu}=(\delta\ro+\delta p)u^0_\mu u^0_\nu-\delta p\bar g^0_{\mu\nu}\eqno(2.29)$$
where $\bar g^0_{\mu\nu}$ is the background (2.1) in comoving coordinates. We have the following diagonal tensor:
$$\delta t_{\mu\nu}=\diag(\delta\ro,\delta pX^2(t),R^2(t)\delta p, R^2\sin^2\te\delta p).\eqno(2.30)$$
This tensor must also be transformed into the Schwarzschild coordinates according to
$$\delta T_{\al\beta}=\delta t_{\mu\nu}{\d\bar x^\mu\over\d \al}{\d\bar x^\nu\over\d \beta}\eqno(2.31)$$
where the bar-coordinates are the comoving ones $(t,r)$. Using the values of the derivatives (2.19) we obtain
$$\delta T_{\al\beta}=\diag(\delta p,\delta\ro{R\over T_L-R},R^2\delta p,R^2\sin^2\te\delta p)\eqno(2.32)$$
It is essential to notice that density and pressure perturbations are interchanged as a consequence of (2.19). The diagonal form shows that only the tensor harmonics $a_{LM}^{(0)}$, $a_{LM}$ and $g_{LM}$ in (2.26-27) contribute in the perturbation theory next section.

\section{Zerilli's perturbation theory applied to the inner Schwarzschild solution}

In this section we always calculate with Schwarzschild coordinates inside the horizon. The metric perturbations $\delta g_{\mu\nu}$ are usually denoted by $h_{\mu\nu}$. If $G_{\mu\nu}$ is the Einstein tensor we have to solve the perturbed Einstein equation
$$\delta G_{\mu\nu}=8\pi G\delta T_{\mu\nu}\eqno(3.1)$$
where $G$ is Newton's constant. Expanding the perturbations and the right side of (3.1) in tensor harmonics, the angular variables $(\te,\phi)$ in the linear equation (3.1) are separated and we are left with 10 linear partial differential equations in $(S,R)$. The 3 axial, odd-parity equations decouple from the 7 polar, even-parity equations. Since the density and pressure perturbations (2.32) only contribute to the polar sector, we restrict to the 7 polar tensor harmonics in the following. A further simplification is obtained by choosing a convenient gauge. In the so-called Regge-Wheeler gauge the 4 tensor harmonics (2.26-27) contribute only ([4], Table I). Then for $L\ge 2$ the metric perturbation is of the form ([4], equ. before (D5))
$$h_{\mu\nu}=\pmatrix{{R-T_L\over R}H_0&H_1&0&0\cr H_1&{R\over R-T_L}H_2&0&0\cr 0&0&R^2K&0\cr 0&0&0&R^2K\sin^2\te\cr}Y_L^M(\te.\phi)\eqno(3.2)$$
where the functions in the matrix depend on $S$ and $R$ only.

Now from the non-diagonal elements of equation (3.1) one obtains the following four homogeneous partial differential equations
([4], equ.(C7a-f)):
$${\d\over\d S}\B({\d K\over\d R}+{1\over R}(K-H_2)-{T_L\over 2R(R-T_L)}K\B)-{L(L+1)\over 2R^2}H_1=0\eqno(3.3)$$
$${\d\over\d R}\B[\B(1-{T_L\over R}\B)H_1\B]-{\d\over\d S}(H_2+K)=0\eqno(3.4)$$
$$-{\d H_1\over \d S}+\B(1-{T_L\over R}\B){\d\over\d R}(H_0-K)+{T_L\over R^2}H_0+{1-T_L/2R\over R}(H_2-H_0)=0\eqno(3.5)$$
$${1\over 2}(H_0-H_2)=0.\eqno(3.6)$$
The remaining diagonal elements yield the following three equations with matter contributions:
$$\B(1-{T_L\over R}\B)^2{\d^2K\over\d R^2}+\B(1-{T_L\over R}\B)\B(3-{5T_L\over 2R}\B){1\over R}{\d K\over\d R}-\B(1-{T_L\over R}\B)^2{1\over R}{\d H_2\over\d R}-$$
$$-\B(1-{T_L\over R}\B){1\over R^2}(H_2-K)-\B(1-{T_L\over R}\B){L(L+1)\over 2R^2}(H_2+K)=8\pi G\delta p\eqno(3.7)$$
$$\B({R\over R-T_L}\B)^2{\d^2K\over\d S^2}-{1-T_L/2R\over R-T_L}{\d K\over\d R}-{2\over R-T_L}{\d H_1\over\d S}+{1\over R}{\d H_0\over\d R}+$$
$$+{1\over R(R-T_L)}(H_2-K)+{L(L+1)\over 2R(R-T_L)}(K-H_0)=8\pi G\delta\ro{R\over T_L-R}\eqno(3.8)$$
$${R\over T_L-R}{\d^2K\over\d S^2}+\B(1-{T_L\over R}\B){\d^2K\over\d R^2}+\B(1-{T_L\over 2R}\B){2\over R}{\d K\over\d R}+{R\over T_L-R}{\d^2H_2\over\d S^2}+$$
$$+2{\d^2H_1\over \d R\d S}-\B(1-{T_L\over R}\B){\d^2H_0\over\d R^2}+{2\over R-T_L}\B(1-{T_L\over 2R}\B){\d H_1\over\d S}-$$
$$-{1\over R}\B(1-{T_L\over 2R}\B){\d H_2\over\d R}-{1\over R}\B(1+{T_L\over 2R}\B){\d H_0\over\d R}-{L(L+1)\over 2R^2}(H_2-H_0)=8\pi GR^2\delta p\eqno(3.9)$$

We have 7 equations for the 4 unknown functions $H_0$, $H_1$, $H_2$ and $K$ and for $\delta\ro$ and $\delta p$. Since $S$ only appears in derivatives we go over to Fourier transformed quantities
$$\hat f(q,R)=(2\pi)^{-1/2}\int f(S,R)e^{iqS}dS,\eqno(3.10)$$ 
which are denoted by the same symbols without hat in the following. Equation (3.6) allows to eliminate $H_0$. Next Zerilli solves (3.3-5)  for the derivatives of $H_1$, $H_2$ and $K$ (his equ.(F3a-c)):
$${d K\over d R}=\B(1-{3T_L\over 2R}\B){1\over T_L-R}K+{1\over R}H_2-{L(L+1)\over 2iq R^2}H_1\eqno(3.11)$$ 
$${d H_2\over d R}=\B(1-{3T_L\over 2R}\B){1\over T_L-R}K+\B(1-{2T_L\over R}\B){1\over R-T_L}H_2-\B[{iq R\over R-T_L}
+{L(L+1)\over 2iq R^2}\B]H_1\eqno(3.12)$$
$${d H_1\over d R}={iq R\over T_L-R}(K+H_2)+{T_L\over R}{1\over T_L-R}H_1.\eqno(3.13)$$
If one substitutes these into (3.7), the left-hand side is zero. The same conclusion follows from (3.9). Consequently, the pressure perturbation $\delta p$ vanishes. But from (3.8) we see that the density perturbation is different from zero equal to ([4] equ.(F4))
$$16\pi GR^2\delta\ro=\B[{3T_L\over R}+(L-1)(L+2)\B]H_2-\B[(L-1)(L+2)+{2q^2R^3\over T_L-R}+\B(1-{3T_L\over 2R}\B){T_L\over R-T_L}\B]K-$$
$$-\B[2iq R+L(L+1){T_L\over 2iq R^2}\B]H_1.\eqno(3.14)$$
Here the evolution equations (3.11-13) have been inserted.
There remains to solve the equations (3.11-13) for $H_1$, $H_2$ and $K$.

\section{Evolution of the perturbations}

From now on we deviate from Zerilli. Zerilli has combined the three first-order equations (3.11-13) with (3.14) to get one second-order equation for one single function. But this second-order equation cannot be reduced to quadratures. Instead we shall solve the linear system (3.11-13) in suitable variables and use the condition (3.14) to select the physical solution for the metric perturbations and to determine the density.

First let us study the density perturbation $\delta\ro$. We consider energy-momentum conservation in the unperturbed metric
$$\nabla^\mu t_{\mu\nu}=0.\eqno(4.1)$$
Using (2.28) and linearizing about the vacuum background we have
$$\delta\dot\ro+\delta\ro\B({\dot X\over X}+2{\dot R\over R}\B)=0\eqno(4.2)$$
because the pressure vanishes. Here the covariant derivative in (4.1) was carried out in comoving coordinates by means of the Christoffel symbols [1], the dot means partial derivative with respect to comoving time. This gives
$${\delta\dot\ro\over\delta\ro}=-{\d\over\d t}\log(\vert X\vert R^2²)$$
which can immediately be integrated
$$\delta\ro(t,r)={\ro_0(r)\over\vert X\vert R^2}\eqno(4.3)$$
where $\ro_0(r)$ is an arbitrary function of the comoving radial coordinate. With our results (2.2) and (2.7) with $Q=1$ we obtain
$$\delta\ro=T_L^2{\ro_0(r)\over\vert\sin^3w\vert\cos w}.\eqno(4.4)$$
This shows that the energy density has a singularity at the Big Bang $w=\pi/2$ and a strong one at the end $w=\pi$. 

Instead of the Schwarzschild variable $R$ we shall use the redshift $z$ (2.8-12) in the following
$$1+z={1\over\al\vert\cot w\vert}\eqno(4.5)$$
or rather
$$x\equiv\al(1+z)=\vert\tan w\vert.\eqno(4.6)$$
We then have
$$X=\vert\cot w\vert={1\over x}\eqno(4.7)$$
$$R=T_L\sin^2w=T_L{x^2\over x^2+1}\eqno(4.8)$$
$$T_L-R={T_L\over x^2+1}\eqno(4.9)$$
and 
$${d\over dR}={(x^2+1)^2\over 2T_Lx}{d\over dx}.\eqno(4.10)$$
Now the left-hand side of (3.14) becomes very simple
$$16\pi GR^2\delta\ro=16\pi G{\ro_0\over \vert X\vert}=16\pi G\ro_0x\eqno(4.11)$$
due to (4.3). Consequently it is good enough to solve the evolution equations (3.11-13) by power series in $x$. However, we shall see that (4.11) gives the density $\delta\ro$ to leading order only, because the higher orders  (3.14) are not contained in (4.2).

In the new variable $x$ our linear system has the following form 
$$K'=-{x\over x^2+1}\B(1+{3\over x^2}\B)K+{2H_2\over x(x^2+1)}-{L(L+1)\over T_Lx^3}H_3\eqno(4.12)$$
$$H'_2=-{x\over x^2+1}\B(1+{3\over x^2}\B)K+{2\over x^2+1}\B(x+{2\over x}\B)H_2-$$
$$-\B[q^2T_L^2{2x^3\over (x^2+1)^2}+{L(L+1)\over x^3}\B]{H_3\over T_L}\eqno(4.13)$$
and
$$H'_3={2T_Lx^3\over (x^2+1)^2}(K+H_2)+{2\over x}H_3.\eqno(4.14)$$
Here the prime is the derivative with respect to $x$ and we have introduced the function
$$H_3={H_1\over iq}\eqno(4.15)$$
in order to have real quantities only. The imaginary nature of $H_1$ comes from the $i$ in the definition (2.27) of the tensor harmonic $a_{LM}^{(1)}$.
We also remove $T_L$ by the additional substitution
$$H_4={H_3\over T_L}={H_1\over iqT_L}.\eqno(4.16)$$

By multiplying the three equations by suitable factors we obtain the following equations without fractions
$$(x^5+x^3)K'=-(x^4+3x^2)K+2x^2H_2-L(L+1)(x^2+1)H_4\eqno(4.17)$$
$$(x^7+2x^5+x^3)H'_2=-(x^6+4x^4+3x^2)K+(2x^6+6x^4+4x^2)H_2-[Q^2x^6+L(L+1)(x^4+2x^2+1)]H_4\eqno(4.18)$$
$$(x^5+2x^3+x)H'_4=2x^4(K+H_2)+2(x^4+2x^2+1)H_4.\eqno(4.19)$$
Here we have introduced the abbreviation
$$Q^2=2q^2T_L^2.\eqno(4.20)$$
The equation (3.14) for the density now reads as follows
$$16\pi GR^2\delta\ro=H_2\B(3+{3\over x^2}+(L-1)(L+2)\B)-K\B[(L-1)(L+2)-2+{Q^2x^6\over (x^2+1)^2}+(x^2+1)\B({1\over 2}+{3\over 2x^2}\B)\B]-$$
$$+H_4\B[{Q^2x^2\over x^2+1}-L(L+1){(x^2+1)^2\over x^4}\B].\eqno(4.21)$$

The system (4.17-19) has three linear independent fundamental solutions and the general solution is a linear combination of them.
Since the perturbations are assumed to be small we must find the bounded solutions. A glance to (4.21) shows that these physical solutions have the following power series
$$K={a_1\over x}+{a_2\over x^3}+{a_3\over x^5}+\ldots\eqno(4.22)$$
$$H_2={b_1\over x}+{b_2\over x^3}+\ldots\quad H_4={c_1\over x}+{c_2\over x^3}+\ldots\eqno(4.23)$$
This shows that the metric perturbations go to zero if the Big Bang $z\to\infty$ or $x\to\infty$ is approached. Substituting this into (4.17-19) and comparing the coefficients we obtain
$$a_2=a_1\B(2-4{L(L+1)+3\over 2Q^2+9}\B)\eqno(4.24)$$
$$b_1=a_1\B({12\over 2Q^2+9}-1\B)\eqno(4.25)$$
$$b_2=a_1\B(10{L(L+1)+21\over 2Q^2+25}-6{L(L+1)+9\over 2Q^2+9}-2\B)\eqno(4.26)$$
$$c_1=-{8a_1\over 2Q^2+9}\eqno(4.27)$$
$$c_2=4a_1\B({L(L+1)+9\over 2Q^2+9}-{L(L+1)+21\over 2Q^2+25}\B).\eqno(4.28)$$
Here $a_1$ is a free overall factor, all other coefficients are fixed. The factor $a_1$ can also be fixed by substituting into (4.21). From the leading order $O(x)$ we find
$$16\pi G\ro_0=-a_1\B(Q^2+{1\over 2}\B)\eqno(4.29)$$ 
so that $a_1(q)$ is determined by the $q$-dependence of the initial energy density $\ro_0(q)$. From (4.21) we then get the higher order perturbations of the density
$$16\pi GR^2\delta\ro=-a_1x\B(Q^2+{1\over 2}\B)+{1\over x}\B[-a_2\B(Q^2+{1\over 2}\B)-a_1(L^2+L-2-2Q^2)+b_1(L^2+L+1)+$$
$$+c_1\B(Q^2-{L(L+1)\over 2}\B)\B]+O\B({1\over x^3}\B).\eqno(4.30)$$
The $L$-dependence shows that the density is essentially anisotropic. However, there is no explicit $M$-dependence. That means a deviation from axial symmetry can only come from an $M$-dependence of the initial condition $a_1$ (primary anisotropies).

\section{Discussion}

The most interesting feature of our first order perturbative solution is that no finite initial condition is needed. The vanishing of the metric perturbations at the Big Bang $x=\infty$ uniquely fixes the solution. On the other hand the energy density becomes singular at the Big Bang due to the first term in (4.30). Then first order perturbation theory is not good enough, we have to go to second order. This is also necessary for calculating the pressure which vanishes in first order. For late times pressure-less anisotropic dust is a good approximation, but the early Universe seems to be much more complicated. 

In standard cosmology one introduces the critical density
$$\ro_\crit={3H_0^2\over 8\pi G}.\eqno(5.1)$$
Then the factor on the left-hand side of (4.30) assumes the following form
$$16\pi GR^2={3\over 2\ro_\crit}{(\al^2+1)^4\over\al^6}\B({x^2\over x^2+1}\B)^2\eqno(5.2)$$
so that equ.(4.30) can be written as
$${\delta\ro\over\ro_\crit}={2\over 3}{\al^6\over (\al^2+1)^4}{(x^2+1)^2\over x^3}\B)\vert a_1\vert\B[Q^2+{1\over 2}+O(x^{-2})\B].\eqno(5.3)$$
The factor $\vert a_1\vert (Q)$ (4.29) determines the density fluctuations at the time of last scattering (Sachs-Wolfe effect). There is no reason to expect $\delta\ro$ near $\ro_\crit$. Since $\delta\ro$ in (3.14) is a Fourier transformed quantity depending on redshift $x=\al(1+z)$, it is directly related to the power spectral function $P(Q)$ which can be determined from galaxy surveys. This will be discussed in a later paper.

The metric perturbations calculated in sect.4 are the basis for computing the CMB anisotropies. In the Gerlach and Sengupta's notation [6] we have to identify
$$h_{00}=-{1\over x^2}H_{2LM},\quad h_{01}=H_{1LM}$$
$$h_{11}=-x^2H_{2LM},\quad r^2K=T_L^2\B({x^2\over x^2+1}\B)^2K_{LM}.\eqno(5.4)$$
Then the integrated Sachs-Wolfe effect can be calculated by a formula due to Tomita [7]. This will be done in a later paper. Finally  we should answer the question what is the small parameter which measures the smallness of $h_{\mu\nu}$ - a basic ingredient of every perturbation theory. As such one can take the inverse redshift $1/z$ or $1/x=X$.

\section{Introduction of Part 2}

Nonstandard cosmology is based on the following two observational facts: (i) the isotropy of the cosmic microwave background and (ii) the smallness of the density of visible matter. The first implies that the cosmic gravitational field is spherically symmetric. According to (ii) we neglect material sources in the zeroth approximation and introduce them by first order perturbation theory. So we start from a spherically symmetric vacuum solution of Einstein's equation.  By Birkhoff's theorem this must be the inner Schwarzschild solution in 
Lema\^itre coordinates with a line element of the form [3]
$$ds^2=dt^2-X(t)^2dr^2-R(t)^2(d\te^2+\sin^2\te d\phi^2).\eqno(6.1)$$
The same solution is singled out of the much larger class of spherically symmetric dust solutions (LTB model) by a theoretical argument [1].
In this paper the free time-dependent functions $X(t)$ and $R(t)$  have been determined in sect.7 in such a way that the measured Hubble diagram is reproduced for redshifts between $0<z<10$.

The next step is the analysis of anisotropic perturbations of the background (6.1). This was started in sect.1-5 by using the results of Zerilli on the perturbation of the outer Schwazschild solution [4]. The first order of perturbation theory leads to a finite matter density, but the pressure remains zero. This was somewhat surprising because the anisotropic perturbations of the gravitational field may give rise to nonzero pressure. So we have decided to reconsider the highly important problem from scratch. Since the library of Princeton University does not deliver a copy of Zerilli's PhD thesis and in [4] no details are given, we have to repeat all calculations carefully. But this is a pleasure because we get an insight into the structure and the evolution of our Universe, a motivation which Zerilli and the later experts on stellar collaps [5] did not have.

In nonstandard cosmology the energy-momentum tensor $t_{\mu\nu}$ in Einstein's equation is obtained as follows. In comoving coordinates (6.1)
this tensor has diagonal elements only. This can be taken as the definition of the comoving coordinates, because in the vacuum background there is no direct definition. Then setting the non-diagonal elements of the perturbed Einstein equations equal to zero gives the perturbations $h_{\mu \nu}$ of the metric. Using these results in the diagonal elements  yields the tensor $t_{\mu\nu}$.

Part 2 is organized as follows. In the next section we give a synopsis of the perturbation calculation. In sect.8  the evolution equations for the metric perturbations are derived. In sect.9 the energy-momentum tensor is calculated, in particular the energy density. In the last section we discuss the application of the results to observations.

\section{The nonstandard background and its perturbations}

We work with comoving coordinates $t,r,\te,\phi$ and choose units so that $c=1=G$. The components of the nonstandard metric tensor are equal to
$$g_{00}=1,\quad g_{11}=-X^2(t),\quad g_{22}=-R^2(t),\quad g_{33}=-R^2(t)\sin^2\te,\eqno(7.1)$$
and zero otherwise. The non-vanishing Christoffel symbols are given by
$$\Gamma^0_{11}=\dot XX,\quad \Gamma^0_{22}=\dot RR,\quad\Gamma^0_{33}=\dot RR\sin²\te\eqno(7.2)$$
$$\Gamma^1_{01}={\dot X\over X},\quad \Gamma^2_{02}={\dot R\over R}=\Gamma^3_{03}\eqno(7.3)$$
$$\Gamma^2_{33}=-\sin\te\cos\te,\quad \Gamma^3_{23}=\cot\te.\eqno(7.4)$$
Here the dot always stands for $\d/\d t$. We also need the Riemann curvature tensor which has the following non-vanishing components
$$R^0_{101}=\ddot XX,\quad R^0_{202}=\ddot RR,\quad R^0_{303}=\ddot RR\sin^2\te\eqno(7.5)$$
$$R^1_{212}=\dot RR{\dot X\over X},\quad R^1_{313}=\dot RR{\dot X\over X}\sin^2\te \quad R^3_{232}=1+\dot R^2.\eqno(7.6)$$

The time dependence of $X(t)$ and $R(t)$ is most easily given in parametric form
$$t=T_L(w-\sin w\cos w)\eqno(7.7)$$
$$X(t)=\cot w,\quad R(t)=T_L\sin^2w,\eqno(7.8)$$
where $T_L$ is of the order of the Hubble time. The Big Bang corresponds to $w=\pi/2$ and $w=\pi$ gives the end of the comoving time when the Universe is infinitely expanded [2]. In the following most calculations are done without using this time dependence, but the relations
$$\dot R(t)=X(t)\quad X^2={T_L\over R}-1\eqno(7.9)$$
will sometimes be used.

It is our aim to solve the linear perturbation equation of Einstein's equation which is of the form
$$\delta G_{\mu\nu}(g_{\al\beta})(h_{\ro\sigma})=8\pi t_{\mu\nu}.\eqno(7.10)$$
Here $h_{\ro\sigma}$ are the metric perturbations of the nonstandard background to be calculated. Since the background is a vacuum solution we have written no delta on the right-hand side, that means the total energy-momentum tensor must come out from (7.10). It is well known that the linear perturbation of the Ricci tensor is obtained by covariant differentiation
$$\delta R_{\mu\nu}=-{1\over 2}\B( \n^\al\n_\al h_{\mu\nu}-\n^\al\n_\nu h_{\mu\al}-\n^\al\n_\mu h_{\nu\al}+\n_\nu\n_\mu h^\al_\al\B).\eqno(7.11)$$
Then the left-hand side of (7.10) becomes
$$\delta R_{\mu\nu}-{1\over 2}g_{\mu\nu}\B( \n^\beta\n^\al h_{\al\beta}-\n_\beta\n^\beta h_\al^\al-h_{\al\beta}R^{\al\beta}\B)-{1\over 2}h_{\mu\nu}R. $$
Following Zerilli we commute the covariant derivatives by means of the curvature tensor which gives
$$2\delta G_{\mu\nu}=-\n^\al\n_\al h_{\mu\nu}+\n_\nu f_\mu+\n_\mu f_\nu-2R^\beta_{\nu\al\nu}h^\al_\beta-\n_\nu\n_\mu h^\al_\al
+R^\beta_\nu h_{\mu\beta}+R^\beta_\mu h_{\nu\beta}+$$
$$-g_{\mu\nu}(\n^\beta f_\beta-\n^\beta\n_\beta h^\al_\al)-Rh_{\mu\nu}+g_{\mu\nu}h_{\al\beta}R^{\al\beta}$$
where
$$f_\mu=\n^\al h_{\mu\al}.\eqno(7.12)$$
This greatly simplifies the calculations because the Ricci tensor vanishes since the background is a vacuum solution. There remains to calculate
$$2\delta G_{\mu\nu}=-\n^\al\n_\al h_{\mu\nu}+\n_\nu f_\mu+\n_\mu f_\nu-2R^\beta_{\nu\al\nu}h^\al_\beta-\n_\nu\n_\mu h^\al_\al
+$$
$$-g_{\mu\nu}(\n^\al f_\al-\n^\beta\n_\beta h^\al_\al).\eqno(7.13)$$

It is necessary to choose a gauge. There is a preferred gauge due to Regge and Wheeler [8] where $h_{\mu\nu}$ is of the form
$$h_{\mu\nu}=\pmatrix{-H_2&XH_1&0&0\cr XH_1&-X^2H_0&0&0\cr 0&0&R^2K&0\cr 0&0&0&R^2K\sin^2\te\cr}Y_l^m(\te.\phi).\eqno(7.14)$$
Here $Y_l^m$ denote the spherical harmonics, the functions $H_0, H_1, H_2$ and $K$ depend on $t$ and $r$ only. Our notation is such that we can compare our results with Zerilli's. It is known that the calculation in the Regge-Wheeler gauge is equivalent with the use of gauge invariant quantities [5]. Nevertheless, as we shall see, it is highly nontrivial that the angular dependence can be separated by means of the ansatz (7.14).  Using the Christoffel symbols (7.2-4) all covariant derivatives can now be calculated in terms of the functions in (7.14). First we calculate the components of the vector $f_\mu$ (7.12):
$$f_0=\B[-\d_0H_2-{1\over X}\d_1H_1-{\dot X\over X}H_0-\B({\dot X\over X}+2{\dot R\over R}\B)H_2+2{\dot R\over R}K\B]Y$$
$$f_1=\B[X\d_0H_1+\d_1H_0+2\B(\dot X+X{\dot R\over R}\B)H_1\B]Y\eqno(7.15)$$
$$f_2=-K\d_2Y,\quad f_3=-K\d_3Y.$$
Next in calculating $\n^\al f_\al$ angular derivatives appear which operate on the spherical harmonics. All second derivatives combine to the square of the angular momentum operator which satisfies
$$\d_2^2Y_l^m+\cot\te\d_2Y^m_l+{1\over\sin^2\te}\d_3^2Y_l^m=-l(l+1)Y_l^m.\eqno(7.16)$$
Details are given in the Appendix.

\section{Evolution of the metric perturbations}

In this section we calculate the non-diagonal elements of the Einstein tensor and put them equal to zero. The simplest computation is $\delta G_{23}$. Here only the terms
$$-\n_3\n_2h_\al^\al=(-\d_3\d_2+\cot\te\d_3)(H_0-H_2-2K)Y\eqno(8.1)$$
and $\n_3f_2$, $\n_2f_3$ contribute in (7.13), so that we obtain
$$2\delta G_{23}=(\cot\te\d_3-\d_3\d_2)(H_0-H_2)Y=0.$$
This yields the first relation
$$H_0(t,r)=H_2(t,r)\eqno(8.2)$$
which has to be used at various places in the following in order to separate the angular dependence.

Next we turn to $\delta G_{02}$. Here we need
$$\n^\al\n_\al h_{02}=2{\dot R\over R}(K-H_2)\d_2Y\eqno(8.3)$$
and
$$\n_2\n_0 h^\al_\al=\B(\d_0-{\dot R\over R}\B)(H_0-H_2-2K)\d_2Y.$$
Then we get
$$2\delta G_{02}=\B[\d_0(K-H_0)-{1\over X}\d_1H_1+H_0\B({\dot R\over R}-{\dot X\over X}\B)-H_2\B({\dot R\over R}+{\dot X\over X}\B)
\B]\d_2Y=0\eqno(8.4)$$
The same equation is obtained from $\delta G_{03}$. This is the first of the three first order linear partial differential equations for the metric perturbations $H_1, H_2, K$ which we are going to derive.

To compare our results with Zerilli's we must introduce the Schwarzschild variable $R$ instead of time $t$. This is easily done by the substitutions
$$\d_0=X\d_R,\quad\dot R=X,\quad \dot X=-{T_L\over 2R^2}\eqno(8.5)$$
and
$$X^2={T_L\over R}-1.\eqno(8.6)$$
Then multiplying  (8.4) by $X$ we find
$$\B({T_L\over R}-1\B)\d_R(K-H_0)-\d_1H_1+H_0\B({3T_L\over 2R^2}-{1\over R}\B)+H_2\B({1\over R}-{T_L\over 2R^2}\B)=0.\eqno(8.7)$$
This agrees with Zerilli's equation (C7e), remembering that $T_L=2m$ in the Schwarzschild solution.

The next non-diagonal element is $\delta G_{12}$. Here we have
$$\n^\al\n_\al h_{12}=2{\dot R\over R}XH_1\d_2Y$$
so that
$$2\delta G_{12}=[\d_1(K+H_2)+X\d_0H_1+2\dot XH_1]\d_2Y=0.\eqno(8.8)$$
After substituting (8.5-6) this is Zerilli's equation (C7d)
$$\B({T_L\over R}-1\B)\d_RH_1+\d_1(K+H_2)-{T_L\over R^2}H_1=0=\eqno(8.9)$$
$$=\d_R\B[\B({T_L\over R}-1\B)H_1\B]+\d_1(K+H_2).\eqno(8.10)$$
From $\delta G_{13}$ we get the same equation.

The most complicated non-diagonal element is $\delta G_{01}$. Here second derivatives appear, for example
$$\sq h_{01}=g^{\al\beta}\d_\al\d_\beta h_{01}=$$
$$=\B(\d_0^2-{1\over X^2}\d_1^2+{l(l+1)\over R^2}+{\cot\te\over R^2}\d_2\B)h_{01}\eqno(8.11)$$
where (7.16) has been used. The angular depending term with $\cot\te$ must cancel. The details are shown in the appendix, where we derive the result
$$\n_\al\n^\al h_{01}=X\d_0^2H_1+\B(\dot X+2{\dot R\over R}X\B)\d_0H_1-{1\over X}\d_1^2H_1-$$
$$-2{\dot X\over X}\d_1(H_0+H_2)+{X\over R^2}l(l+1)H_1-\B(4{\dot X^2\over X}+2{\dot R^2\over R^2}X\B)H_1.\eqno(8.12)$$
The expressions for $\n_1f_0$ and $\n_0f_1$ are also given in the appendix. We get a contribution from the Riemann tensor
$$R^1_{001}={\ddot X\over X}$$
which is multiplied by $-2h_1^0=-2XH_1$. The final result is
$${\delta G_{01}\over Y}=\d_0\d_1K-{\dot R\over R}\d_1H_2+\B({\dot R\over R}-{\dot X\over X}\B)\d_1K-X{l(l+1)\over 2R^2}H_1=0.\eqno(8.13)$$
With the substitutions (8.5-6) we have
$$\d_R\d_1K-{1\over R}\d_1H_2+\B({1\over R}+{T_L\over 2R(T_L-R)}\B)\d_1K-{l(l+1)\over 2R^2}H_1=0.\eqno(8.14)$$
This is Zerilli's equation (C7b).

Taking (8.2) into account, we have now obtained the three partial differential equations (8.7), (8.9) and (8.14) for the 3 unknown functions $H_1, H_2$ an $K$ of $(t,r)$. Since the $r$-dependence only appears in derivatives $\d_1$  we go over to Fourier transformed quantities
$$\hat f(R,q)=(2\pi)^{-1/2}\int f(R,r)e^{iqr}dr.\eqno(8.15)$$
Then the derivative $\d_1$ goes over to a factor $-iq$ and we omit the hat in the following. Instead of partial differential equations we now have the following 3 ordinary differential equations in time or $R$:
$$\d_RK={H_2\over R}-{3T_L-2R\over 2R(T_L-R)}K-{l(l+1)\over 2R^2}H_3\eqno(8.16)$$
$$\d_RH_2={2T_L-R\over R(T_L-R)}H_2-{3T_L-2R\over 2R(T_L-R)}K+\B({q^2R\over R-T_L}-{l(l+1)\over 2R^2}\B)H_3\eqno(8.17)$$
$$\d_RH_3={R\over T_L-R}(K+H_2)+{T_L\over R(T_L-R)}H_3.\eqno(8.18)$$
Here we have introduced the function
$$H_3={H_1\over iq}\eqno(8.19)$$
in order to have pure real equations. These equations govern the time evolution of the metric perturbations and are further discussed in the last section.

\section{Determination of the energy-momentum tensor}

Now we consider the diagonal elements of the perturbed Einstein's equation (7.10). Since the metric perturbations are fixed by the results of the last section, the diagonal elements determine the (diagonal) tensor 
$$t_{\mu\nu}={\rm diag} (\ro,-g_{11}p_1,-g_{22}p_2,-g_{33}p_3)\eqno(9.1)$$
on the right-hand side of (7.10). We start with $\delta G_{11}$.
The separate pieces are given in the appendix. The final result 
$${\delta G_{11}\over X^2Y}=\B({T_L\over R}-1\B)\d_R^2K+\B[{3\over R}\B({T_L\over R}-1\B)-{T_L\over 2R^2}\B]\d_RK-$$
$$-\B({T_L\over R}-1\B){1\over R}\d_RH_2+{1\over R^2}(H_2-K)\eqno(9.2)$$
agrees with Zerilli's equation (C7a). By differentiating (8.16) with respect to $R$ and then substituting (8.16) and (8.17-18) a second time into (9.2) we get zero. That means the radial pressure $t_{11}=-g_{11}p_r$ vanishes.

The next diagonal element $\delta G_{22}$ is more complicated. The calculation of $\n_\al\n^\al h_{22}$ is given by (A.7) in the appendix. Next we need
$$\n_2f_2=\d_2f_2-\Gamma_{22}^0f_0=-K\d_2^2Y+$$
$$+\dot RR\B[\d_0H_2+{1\over X}\d_1H_1+{\dot X\over X}(H_0+H_2)-2{\dot R\over R}(K-H_2)\B]Y.\eqno(9.3)$$
This gets multiplied by 2 in (7.12). From the Riemann tensor we get three contributions
$$-2R^0_{202}h_0^0=2\ddot RRH_2Y,\quad -2R^1_{212}h_1^1=-2R\dot R{\dot X\over X}H_0Y$$
$$-2R^3_{232}h_3^3=2(1+\dot R^2)KY.\eqno(9.4)$$
The large last contribution in (7.13) with $-g_{22}$ is taken from (A.4). Putting everything together we finally obtain
$${2\delta G_{22}\over R^2Y}=\d_0^2K-{1\over X^2}\d_1^2K-\d_0^2H_0-{1\over X^2}\d_1^2H_2-{2\over X}\d_0\d_1H_1+$$
$$+\d_0K\B(2{\dot R\over R}+{\dot X\over X}\B)-\d_0H_2\B({\dot R\over R}+{\dot X\over X}\B)-2\d_1H_1\B({\dot R\over RX}+{\dot X\over X^2}\B)-$$
$$-\d_0H_0\B({\dot R\over R}+2{\dot X\over X}\B)-(H_2+H_0)\B({\ddot X\over X}+2{\dot X\dot R\over XR}\B)+{l(l+1)\over R^2}(H_2-H_0)+$$
$$+2K\B({1\over R^2}+{\ddot R\over R}+{\dot X\dot R\over XR}\B).\eqno(9.5)$$

After substituting $\d_0$ by $\d_R$ we obtain Zerilli's equation (C7f) because the last term in (9.5) vanishes by (7.9): 
$$\B({T_L\over R}-1\B)\d_R^2(K-H_0)-{R\over T_L-R}\d_1^2(K+H_2)-2\d_R\d_1H_1+$$
$$+{2\over R}\B({T_L\over2R}-1\B)\d_RK+\B({1\over R}+{T_L\over 2R^2}\B)\d_RH_0+\B({1\over R}-{T_L\over 2R^2}\B)\d_RH_2+$$
$$+{2\over T_L-R}\B(1-{T_L\over 2R}\B)\d_1H_1={\delta G_{22}\over R^2Y}.\eqno(9.6)$$
Again we substitute all derivatives by the evolution equations (8.16-18). Since we get zero again, the second pressure components $t_{22}$ also vanishes.

In $\delta G_{33}$ the angular dependence is most complicated. Zerilli probably did not calculate it because we get some new results.  We have
$$\n_\al\n^\al h_{33}=\sin^2\te\B(R^2\d_0^2K+(2\dot RR+R^2{\dot X\over X})\d_0K-{R^2\over X^2}\d_1^2K\B)Y+$$
$$+K\B[\sin^2\te\B(l(l+1)-2\dot R^2\B)+4\cos^2\te-2+4\sin\te\cos\te{\d_2Y\over Y}\B]Y+$$
$$+2H_2\dot R^2\sin^2\te Y.\eqno(9.7)$$
In 
$$\n_3f_3=-\d_3^2KY-\Gamma_{33}^\al f_\al\eqno(9.8)$$
there is even a break of axial symmetry due to the relation
$$\d_3^2Y_l^m(\te,\phi)=-m^2Y_l^m.\eqno(9.9)$$
We find
$$\n_3 f_3=m^2KY-\sin\te\cos\te K\d_2Y-$$ 
$$+\dot RR\sin^2\te \B(\d_0H_2+{1\over X}\d_1H_1+{\dot X\over X}H_0+({\dot X\over X}+2{\dot R\over R})H_2-2{\dot R\over R}K\B)Y.\eqno(9.10)$$
There are three terms from the Riemann tensor $R^0_{303}$, $R_{313}^1$ and $R^2_{323}$ and
$$\n_3\n_3h_\al^\al=\d_3²h_\al^\al-\Gamma_{33}^\al\d_\al h_\beta^\beta=-m^2(H_0-H_2-2K)Y-$$
$$-\dot RR\sin^2\te\d_0(H_0-H_2-2K)Y+\sin\te\cos\te (H_0-H_2-2K)\d_2Y.\eqno(9.11)$$
The last big piece in (7.13) follows from (A.4) in the appendix.

The final result is
$${2\delta G_{33}\over R^2\sin^2\te Y}=\d_0^2(K-H_0)-{1\over X^2}\d_1^2(K+H_2)-{2\over X}\d_0\d_1H_1+$$
$$+\d_0K\B(2{\dot R\over R}+{\dot X\over X}\B)-\d_0H_2\B({\dot R\over R}+{\dot X\over X}\B)-\d_0H_0\B({\dot R\over R}+2{\dot X\over X}\B)-
2\d_1H_1\B({\dot R\over RX}+{\dot X\over X^2}\B)+$$
$$+2K\B[1+{\ddot R\over R}+{\dot R^2\over R^2}+{\dot X\over X}{\dot R\over R}\B]+{2K\over R^2\sin^2\te}\B(1-2\cos^2\te
-2\sin\te\cos\te{\d_2Y\over Y}\B).\eqno(9.12)$$
As far as the terms with derivatives are concerned, this agrees with (9.5); we have left out the terms with $H_0=H_2$ which have zero factors after substituting (8.5-6). It follows from the previous analysis of $\delta G_{22}$ that these derivative terms vanish if the evolution equations (8.16-18) are taken into account. Then we conclude
$$\delta G_{33}=K\sin^2\te[1+\ddot RR+\dot R^2+{\dot X\over X}\dot RR]Y+$$
$$+K(1-2\cos^2\te)Y-2K\sin\te\cos\te\d_2Y.\eqno(9.13)$$
The square bracket vanishes in virtue of (7.9), but the remaining two terms survive.
This gives a non-zero pressure $p_3=p_\phi$. It also shows that the separation of the angular variables is no longer complete.
The two pressure terms come from the square bracket in (9.7) and are not compensated by other contributions.

The most interesting diagonal element is $\delta G_{00}$ which determines the matter density $\ro$ (9.1).  The long calculation of $\n_\al\n^\al h_{00}$ is again shown in the appendix (A.6). Since there is no Christoffel symbol with $\Gamma_{00}$ we simply have
$$\n_0 f_0=\d_0f_0$$
and
$$\n_0\n_0 h_\al^\al=\d_0^2(H_0-H_2-2K).$$
The Riemann tensor gives contributions from $R_{010}^1$, $R_{020}^2$ and $R_{030}^3$. The last big piece is taken from (A.4) again.           Collecting all terms we arrive at
$${2\delta G_{00}\over Y}={2\over X^2}\d_1^2K+4{\dot R\over RX}\d_1H_1+2{\dot R\over R}\d_0H_0-2\B({\dot R\over R}+{\dot X\over X}\B)\d_0K+$$
$$+H_2\B(2{\dot R^2\over R^2}-2{\ddot R\over R}-{\ddot X\over X}+4{\dot X\dot R\over XR}\B)+H_0\B({\ddot X\over X}+2{\dot X\dot R\over XR}\B)-$$
$$-2K\B({\ddot R\over R}+{\dot R^2\over R^2}+{\dot X\dot R\over XR}\B)+{l(l+1)\over R^2}(H_0-K).\eqno(9.14)$$
After substituting Schwarzschild variables
$$\B({T_L\over R}-1\B)^{-1}\d_1^2K+{2\over R}\d_1H_1+{1\over R}\B({T_L\over R}-1\B)\d_R(H_0-K)+$$
$$+{T_L\over 2R²}\d_RK+{1\over R^2}(K-H_2)+{l(l+1)\over 2R^2}(H_0-K)={\delta G_{00}\over Y}\eqno(9.15)$$
this agrees with Zerilli's equation (C7c).

According to (7.19) and (9.1) the result (9.15) gives the matter density $8\pi\ro/Y$. After substituting $\d_1$ by $-iq$ and eliminating the derivatives $\d_R$ with help of the evolution equations (8.16-18) we obtain
$${8\pi\ro\over Y}=K\B[{1\over R^2}-{l(l+1)\over 2R^2}-{3T_L\over 4R^3}-{q^2R\over T_L-R}-{T_L\over 4R^2(T_L-R)}\B]+$$
$$+H_2\B[{3T_L\over 2R^3}-{1\over R^2}+{l(l+1)\over 2R^2}\B]+H_3\B[{q^2\over R}-{T_L\over 4R^4}l(l+1)\B].\eqno(9.16)$$
Due to the $l$-dependence this is an anisotropic matter density which will be discussed in the next section.

\section{Discussion}

To apply the perturbative results to our Universe we must first integrate the evolution equations. This is best done by using the variable
$$x=\al(1+z)\eqno.(10.1)$$
which is directly related to the redshift $z$. The parameter $\al$ is determined by the Hubble diagram which gives the value $\al=2.59$. The transformation of variables is given by
$$R=T_L{x^2\over x^2+1},\quad X={1\over x}\eqno.(10.2)$$
$$T_L-R={T_L\over x^2+1}\eqno.(10.3)$$
and 
$${d\over dR}={(x^2+1)^2\over 2T_Lx}{d\over dx}.\eqno.(10.4)$$
In the new variable $x$ our the evolution equations read
$$K'=-{x\over x^2+1}\B(1+{3\over x^2}\B)K+{2H_2\over x(x^2+1)}-{l(l+1)\over T_Lx^3}H_3\eqno.(10.5)$$
$$H'_2=-{x\over x^2+1}\B(1+{3\over x^2}\B)K+{2\over x^2+1}\B(x+{2\over x}\B)H_2-$$
$$-\B[q^2T_L^2{2x^3\over (x^2+1)^2}+{l(l+1)\over x^3}\B]{H_3\over T_L}\eqno.(10.6)$$
and
$$H'_3={2T_Lx^3\over (x^2+1)^2}(K+H_2)+{2\over x}H_3.\eqno.(10.7)$$
Here the prime is the derivative with respect to $x$.

The system of 3 linear first order ode's can be numerically integrated by any ode-solver, for example DSolve of Mathematica. For not to small $x$ a power series expansion can be used. As shown in sect.4 it is of the form
$$K={a_1\over x}+{a_2\over x^3}+{a_3\over x^5}+\ldots\eqno.(10.8)$$
$$H_2={b_1\over x}+{b_2\over x^3}+\ldots\quad H_4={H_3\over T_L}={c_1\over x}+{c_2\over x^3}+\ldots\eqno.(10.9)$$
Here all coefficients are determined by $a_1$ which fixes the overall normalization of the metric perturbations. A stable numerical integration of the evolution equations must go from large to small redshifts. So one starts at $z=10$, say, with normalization $a_1=1$ calculating the initial values by the power series .(10.8-9). Then one integrates down to the present $z=0$ where one can compare the result with measured matter density (9.16). This then gives the correct normalization. It turns out that the numerical solution of the evolution equations is only necessary for the late Universe $z<10$, for $z>10$ the power series .(10.8-9) can be used if $l$ is not too big.

For the approach to the Big Bang $z=\infty$ the power series are perfect. The metric perturbation go to zero as $1/x$. However the energy density (9.16) grows proportional to $x$. In fact we have shown in [3] eq..(10.3) that
$${\ro\over\ro_{\rm crit}}={2\over 3}{\al^6\over (\al^2+1)^4}{(x^2+1)^2\over x^3}\B)\vert a_1\vert\B[Q^2+{1\over 2}+O(x^{-2})\B]\eqno.(10.10)$$
where $\ro_{\rm crit}$ is the critical density. It is a nice feature of nonstandard cosmology that the Big Bang is not a singularity of Einstein's equation, because it corresponds to the horizon in the Schwarzschild solution. But the growth of the energy density .(10.10) indicates that first order perturbation theory cannot be the whole story. We shall treat second order perturbation theory in another paper.
The longitudinal pressure $p_\phi$ obtained from (9.13) is proportional to $K=O(x^{-1})$. So it is small compared to the density $\ro=O(x)$
for large redshift.

\section{Appendix}

The computation of the very many covariant derivatives is cumbersome. Here is an example:
$$\n^\al f_\al=g^{\al\beta}\n_\beta f_\al=\d_0f_0-{1\over X^2}(\d_1f_1-\Gamma_{11}^0f_0)-$$
$$-{1\over R^2}(\d_2f_2-\Gamma^0_{22}f_0)-{1\over R^2\sin^2\te}(\d_3f_3-\Gamma_{33}^\al f_\al).$$
Here in the last term two Christofel symbols $\al=0,2$ contribute. Substituting the explicit expressions (7.19) and using (7.16) we finally obtain
$${\n^\al f_\al\over Y}=-\d_0^2H_2-{1\over X^2}\d_1^2H_0-{2\over X}\d_0\d_1H_1-\B(2{\dot X^2\over X^2}+4{\dot R\over RX}\B)\d_1H_1-$$
$$-\B(2{\dot X\over X}+4{\dot R\over R}\B)\d_0H_2-{\dot X\over X}\d_0H_0+2{\dot R\over R}\d_0 K+$$
$$-H_0\B({\ddot X\over X}+2{\dot R\dot X\over RX}\B)-H_2\B({\ddot X\over X}+2{\ddot R\over R}+2{\dot R^2\over R^2}+4{\dot R\dot X\over RX}
\B)+$$
$$+2K\B({\ddot R\over R}+{\dot R^2\over R^2}+{\dot R\dot X\over RX}-{l(l+1)\over 2R^2}\B).\eqno(A.1)$$

Another long calculation is
$$\n_\al\n^\al h_{01}=\sq h_{01}-2\Gamma_{01}^1\d^1h_{11}-2\Gamma_{11}^0\d^1h_{00}-2\Gamma_{01}^1\d^0h_{01}+$$
$$-\B(g^{11}\Gamma_{11}^0\d_0+g^{22}\Gamma_{22}^0\d_0+g^{33}\Gamma_{33}^0\d_0+g^{33}\Gamma_{33}^2\d_2\B)h_{01}+$$
$$+\B(g^{11}\Gamma_{01}^1\Gamma_{11}^0+g^{22}\Gamma_{02}^2\Gamma_{22}^0+g^{33}\Gamma_{03}^3\Gamma_{33}^0\B)h_{01}-h_{01}\d_0\Gamma_{01}^1+$$
$$+\B(g^{11}\Gamma_{11}^0\Gamma_{01}^1+g^{22}\Gamma_{22}^0\Gamma_{01}^1+g^{33}\Gamma_{33}^0\Gamma_{01}^1\B)h_{01}+$$
$$+\B(g^{00}\Gamma_{01}^1\Gamma_{01}^1+g^{11}\Gamma_{11}^0\Gamma_{01}^1\B)h_{01}+2g^{11}\Gamma_{01}^1\Gamma_{11}^0h_{01}.\eqno(A.2)$$
Here the term with $\Gamma_{33}^2$ cancels the angular term with $\cot\te$ in $\sq h_{01}$. After substituting all explicit expressions we arrive at the result (8.12).

For the diagonal elements we need the last term in (7.13):
$$\n^\beta\n_\beta h_\al^\al=g^{\mu\beta}[\d_\beta\d_\mu(H_0-H_2-2K)Y-\Gamma_{\mu\beta}^\al\d_\al(H_0-H_2-2K)Y=$$
$$=\sq(H_0-H_2-2K)Y-\B[-\Gamma_{11}^0{1\over X^2}\d_0-\Gamma_{22}^0{1\over R^2}\d_0-$$
$$-\Gamma_{33}^0{1\over R^2\sin^2\te}\d_0-\Gamma_{33}^2{1\over R^2\sin^2\te}\d_2\B](H_0-H_2-2K)Y=$$
$$=\B[\d_0^2-{1\over X^2}\d_1^2+{l(l+1)\over R^2}+\B({\dot X\over X}+2{\dot R\over R}\B)\d_0\B](H_0-H_2-2K).\eqno(A.3)$$
Combining this with (A.1) gives
$$\n^\al f_\al-\n^\beta\n_\beta h^\al_\al=\d_0^2(2K-H_0)-{1\over X^2}\d_1²(2K-H_2)-{2\over X}\d_0\d_1H_1-$$
$$-\B(2{\dot X\over X^2}+4{\dot R\over XR}\B)\d_1H_1-\B(2{\dot X\over X}+2{\dot R\over R}\B)\d_0H_0-\B({\dot X\over X}+2{\dot R\over R}\B)
\d_0H_2+$$
$$+\B(2{\dot X\over X}+6{\dot R\over R}\B)\d_0K-\B({\ddot X\over X}+2{\dot X\dot R\over XR}\B)H_0
-\B({\ddot X\over X}+2{\ddot R\over R}+2{\dot R^2\over R^2}+4{\dot X\dot R\over XR}\B)H_2+$$
$$+\B(2{\ddot R\over R}+2{\dot R^2\over R^2}+2{\dot X\dot R\over XR}\B)K+{l(l+1)\over R^2}(H_2-H_0+K).\eqno(A.4)$$

For $\delta G_{11}$ we need
$$\n_\al\n^\al h_{11}=\sq h_{11}-4\Gamma_{\al 1}^\ro\d^\al h_{\ro 1}-g^{\al\beta}\Gamma_{\al\beta}^\ro\d_\ro h_{11}+$$
$$+2g^{\al\beta}h_{\ro 1}(-\d_\beta\Gamma_{1\al}^\ro+\Gamma_{\al\beta}^\sigma\Gamma_{\sigma 1}^\ro
+\Gamma_{1\beta}^\sigma\Gamma_{\al\sigma}^\ro)+2g^{\al\beta}h_{\ro\sigma}\Gamma_{\beta 1}^\ro\Gamma_{\al 1}^\sigma$$
$$=-X^2\d_0²H_0+\d_1^2H_0-\B(X\dot X+2{\dot R\over R}X^2\B)\d_0H_0+4\dot X\d_1H_1+$$
$$+4\dot X\d_1H_1+2\dot X^2H_2+H_0\B(2\dot X^2-{X^2\over R^2}l(l+1)\B)\eqno(A.5)$$
and
$$\n_1 f_1=\d_1f_1-\Gamma_{11}^0f_0=\B[\d_1\d_0(XH_1)+\d_1^2H_0+$$
$$+\B(\dot X+2{\dot RX\over R}\B)\d_1H_1+X\dot X\d_0H_2+\dot X\d_1H_1+\dot X^2(H_0+H_2)-$$
$$-2\dot XX{\dot R\over R}(K-H_2)\B]Y.\eqno(A.6)$$
We get contributions from the components $R^0_{101}$, $R_{121}^2$ and $R_{131}^3$ of the Riemann tensor. In addition  (A.4)must be used for the last term in (7.13).

Next we compute
$$\n_\al\n^\al h_{22}=\sq h_{22}-4\Gamma_{02}^2\d^0h_{22}-\B(g^{11}\Gamma_{11}^0\d_0+g^{22}\Gamma_{22}^0\d_0+$$
$$+g^{33}\Gamma_{33}^0\d_0+g^{33}\Gamma_{33}^2\d_2\B)h_{22}-2h_{22}\d_0\Gamma_{20}^2-2h_{22}\B({\dot X\over X}+2{\dot R\over R}\B){\dot R
\over R}+$$
$$+2h_{22}\B(g^{00}\Gamma_{20}^2\Gamma_{02}^2+g^{22}\Gamma_{22}^0\Gamma_{02}^2+g^{33}\Gamma_{23}^3\Gamma_{33}^2\B)+$$
$$+2\B(h_{22}g^{00}(\Gamma_{02}^2)^2+h_{00}g^{22}(\Gamma_{22}^0)^2+h_{33}g^{33}(\Gamma_{32}^3)^2\B)$$
$$=R^2\d_0^2K-{R^2\over X^2}\d_1^2K+\B(2\dot RR+R^2{\dot X\over X}\B)\d_0K+2\dot R^2H_2+[l(l+1)-2\dot R^2]K.\eqno(A.7)$$

For $\delta G_{00}$ we need
$$\n_\al\n^\al h_{00}=\sq h_{00}-4\Gamma_{10}^1\d^1h_{10}-(g^{11}\Gamma_{11}^0+g^{22}\Gamma_{22}^0+g^{33}\Gamma_{33}^0)\d_0h_{00}-$$
$$-g^{33}\Gamma_{33}^2\d_2h_{00}+2h_{00}(g^{11}\Gamma_{01}^1\Gamma_{11}^0+g^{22}\Gamma_{02}^2\Gamma_{22}^0+g^{33}\Gamma_{03}^3\Gamma_{33}^0)+$$
$$+2\B[h_{11}(\Gamma_{10}^1)^2g^{11}+h_{22}(\Gamma_{20}^2)^2g^{22}+h_{33}(\Gamma_{30}^3)^2g^{33}\B]=$$
$$=-\B(\d_0^2-{1\over X^2}\d_1^2+{l(l+1)\over R^2}\B)H_2+4{\dot X\over X^2}\d_1H_1-\B({\dot X\over X}+2{\dot R\over R}\B)\d_0H_2+$$
$$+H_2\B(2{\dot X^2\over X^2}+4{\dot R^2\over R^2}\B)+2{\dot X^2\over X^2}H_0-4{\dot R^2\over R^2}K.\eqno(A.8)$$

\end{document}